\documentclass[aps,showpacs,onecolumn,superscriptaddress]{revtex4}
\usepackage{graphics,color}
\usepackage{epsfig}
\usepackage{dcolumn}
\usepackage{bm}
\usepackage{epsfig}
\usepackage{mathrsfs}
\usepackage{amsfonts}
\usepackage{graphicx}
\usepackage{amsmath}
\usepackage{amssymb}

\begin{document}

\title{ Entanglement and Berry Phase in a $9\times 9$ Yang-Baxter system }
\author{Chunfang Sun}
\address{School of Physics, Northeast Normal University,
Changchun 130024, People's Republic of China}
\author{Kang Xue}
\email{Xuekang@nenu.edu.cn}
\address{School of Physics, Northeast Normal University,
Changchun 130024, People's Republic of China}
\author{Gangcheng Wang}
\address{School of Physics, Northeast Normal University,
Changchun 130024, People's Republic of China}

\begin{abstract}
 A M-matrix which satisfies the Hecke algebraic relations is
presented. Via the Yang-Baxterization approach, we obtain a unitary
solution $\breve{R}(\theta,\varphi_{1},\varphi_{2})$ of Yang-Baxter
Equation. It is shown that any pure two-qutrit entangled states can
be generated via the universal $\breve{R}$-matrix assisted by local
unitary transformations. A Hamiltonian is constructed from the
$\breve{R}$-matrix, and Berry phase of the Yang-Baxter system is
investigated. Specifically, for $ \varphi_{1}=\varphi_{2}$, the
Hamiltonian can be represented based on three sets of SU(2)
operators, and three oscillator Hamiltonians can be obtained. Under
this framework, the Berry phase can be
interpreted.
\end{abstract}

\pacs{03.67.Mn, 02.40.-k,03.65.Vf} \maketitle

\vspace{0.3cm}

\section{Introduction}
In 1984, Berry\cite{berry} predicted that a quantum system acquires
a geometrical phase, in additional to a dynamical phase, if the
environment or the Hamiltonian returns to its initial state
adiabatically. Simon\cite{BS} was the first to recast the
mathematical formalism of berry phase within the elegant language of
differential geometry and fibre bundles. Berry's theory has been
generalized by extending it to non-adiabatic evolution,non-cyclic
and non-unitary evolution, mixed states or non-abelian geometric
phases\cite{aa,Esj,JR,DELC,FA}. A vast number of experiments have
verified its characteristics in various systems, including NMR
\cite{DGRA,MVM}, the closely related technique of NQR
\cite{RT,SGM,JA}, optical systems \cite{RY}, and so on. Recently
more and more works have been attributed to it \cite{AAHQJZ}because
of its possible applications to quantum computation \cite{J.
Jones,W. K.,A. Ekert}. Such interest is motivated by the belief that
geometric quantum gates should exhibit an intrinsic fault tolerance
in the presence of some kind of external noise due to the geometric
nature of the Berry phase. Quantum gate operations can be
implemented through the geometric effects on the wave function of
the systems. On the other hand, entanglement is a bizarre feature of
quantum theory, has been recognized as an important resource for
applications in quantum information and computation processing
\cite{Bennett,ben1,ben2,mdmv}.

Very recently, braiding operators and Yang-Baxter equation (YBE)
\cite{yang,baxter,drin} have been introduced to the field of quantum
information and quantum computation
\cite{qiybe1,kauffman1,qiybe3,zkg,zg,ckg,ckg2,cxg1,S. W. Hu,wang1}.
Ref. \cite{qiybe1} investigated quantum computation by anyons based
on quantum braids. Ref. \cite{kauffman1} has explored the role of
unitary braiding operators in quantum computation. It is shown that
the braid matrix can be identified as the universal quantum gate
\cite{kauffman1,qiybe3,zkg}. This motivates a novel way to study
quantum entanglement and the Berry phase based on the theory of
braiding operators, as well as YBE. The first step along this
direction is initiated by Zhang, Kauffman and Ge \cite{zkg}. In Ref.
\cite{zkg}, the Bell matrix generating two-qubit entangled states
has been recognized to be a unitary braid transformation. Later on,
an approach to describe Greenberger-Horne-Zeilinger (GHZ) states, or
N-qubit entangled states based on the theory of unitary braid
representations has been presented in Ref. \cite{zg}. Chen and his
co-workers \cite{ckg,ckg2} utilized unitary braiding operators to
realize entanglement swapping and generate the GHZ states, as well
as the linear cluster states. With the unitary
$\breve{R}(\theta,\phi)$ matrix, the authors constructed a
Hamiltonian. The Berry phase and quantum criticality of the
Yang-Baxter system have been explored accordingly. In a very recent
work \cite{cxg1}, it is found that any pure two-qudit entangled
state can be achieved by a universal Yang-Baxter Matrix assisted by
local unitary transformations. However, The unitary solution
$\check{R}(x)$ of YBE, which is presented in Ref. \cite{cxg1}, only
depends on a parameter $\theta$ which is time-independent. We can't
explore the evolution of the Yang-Baxter system. Motivated by this,
in this paper we present a unitary solution of YBE,
$\check{R}(\theta,\varphi_{1},\varphi_{2})$, where $\varphi_{1}$ and
$\varphi_{2}$ are time-dependent, while $\theta$ is
time-independent. Thus, we can explore the evolution of the
Yang-Baxter system by constructing a Hamiltonian from the matrix
$\check{R}(\theta,\varphi_{1},\varphi_{2})$. This in turn allows us
to investigate the Berry phase in the entanglement space.

The paper is organized as follows: In Sec \ref{sec2}, we present a
$9 \times 9$ Yang-Baxter matrix
$\breve{R}(\theta,\varphi_{1},\varphi_{2})$. In the following, we
investigate the entanglement properties of it. It is shown that the
arbitrary degree of entanglement for two-qutrit entangled states can
be generated via the unitary $\breve{R}$-matrix acting on the
standard basis. In fact, we can prove that this unitary matrix
$\breve{R}$ is local equivalent to the solution in Ref.\cite{cxg1}.
So we can say that all pure two-qutrit entangled states can be
generated via the universal $\breve{R}$-matrix assisted by local
unitary transformations. In Sec \ref{sec3}, we construct a
Hamiltonian from the unitary $\breve{R}$-matrix. The Berry phase of
this system is investigated. Specifically, for $
\varphi_{1}=\varphi_{2}$, the three nonzero Hamiltonian subsystems
all are shown to be equivalent to oscillator systems of two
fermions. We end with a summary.

\section{Unitary solution of yang-baxter equation and entanglement}\label{sec2}

   As is known, The YBE is given by
\begin{equation}
\breve{R}_{i}(x)\breve{R}_{i+1}(xy)\breve{R}_{i}(y)=\breve{R}_{i+1}(y)\breve{%
R}_{i}(xy)\breve{R}_{i+1}(x) , \label{4}
\end{equation}
where x and y are spectrum parameters. The notation $\breve{R}_{i}(x)\equiv%
\breve{R}_{i,i+1}(x)$ is used, $\breve{R}_{i,i+1}(x)$ implies $1
_{1}\otimes1_{2}\otimes1_{3}\cdots \otimes%
\breve{R}_{i,i+1}(x)\otimes\cdots\otimes1_{n}$, $ 1_{j}$ represents
the unit matrix of the $j$-th particle, and $ x=e^{i\theta}$ is a
parameter related to the degree of entanglement. Let the unitary
Yang-Baxter $\breve{R}$-matrix for two qutrits be the form
\begin{equation}
\breve{R}_{i}(x)=\rho(x)[\mathbf{1}_{i}+G(x)M_{i}], \label{5}
\end{equation}
where $\rho(x)$ and $G(x)$ are some functions needed to determine later on, $%
\mathbf{1}_{i}=1_{i}\otimes1_{i+1}$, and the Hermitian matrices
$M_{i}$'s (i.e., $M_{i}=M_{i}^{+})$ satisfy the Hecke algebraic
relations: $M_{i}M_{i+1}M_{i}+gM_{i}=M_{i+1}M_{i}M_{i+1}+gM_{i+1}$, $%
M_{i}^{2}=\alpha M_{i}+\beta \mathbf{1}_{i}$, with
$\beta=g=2\alpha^{2}$ while $\alpha\neq0$. Substituting Eq.
(\ref{5}) into Eq. (\ref{4}), one has $G(x)+G(y)+\alpha
G(x)G(y)=[1+gG(x)G(y)]G(xy)$. The initial condition
$\breve{R}_{i}(1)=I_{i}$ leads to $G(1)=0$, $\rho(1)=1$. In addition, the unitary condition $\breve{R}_{i}^{+}(x)=%
\breve{R}_{i}^{-1}(x)=\breve{R}_{i}(x^{-1})$ yields
$G(x)+G(x^{-1})+\alpha G(x)G(x^{-1})=0$,
$\rho(x)\rho(x^{-1})[1+\beta G(x)G(x^{-1})]=1$. For convenience, we
restrict ourselves on $\alpha=1$, and $\beta=g=2$. As a result, one
has\newline
$G(x)=-\frac{x-x^{-1}}{2x+x^{-1}}$, $\rho(x)=\frac{%
2x+x^{-1}}{3}$.

In this work, we choose basis $\{|00\rangle, |01\rangle, |02\rangle,
|10\rangle, |11\rangle, |12\rangle, |20\rangle, |21\rangle,
|22\rangle\}$ as the standard basis. The $9\times 9$ matrix M is
realized as,
\begin{eqnarray}
M=\left(
\begin{array}{ccccccccc}
0 & 0 & 0 & 0 & \frac{q_{2}^{2}}{q_{1}^{2}} & 0 & 0 & 0 & q_{2}^{2} \\
0 & 0 & 0 & 0 & 0 & q_{2} & q_{1} & 0 & 0 \\
0 & 0 & 0 & \frac{1}{q_{1}} & 0 & 0 & 0 & \frac{q_{2}}{q_{1}} & 0 \\
0 & 0 & q_{1} & 0 & 0 & 0 & 0 &  q_{2} & 0 \\
\frac{q_{1}^{2}}{q_{2}^{2}} & 0 & 0 & 0 & 0 & 0 & 0 & 0 & q_{1}^{2} \\
0 & \frac{1}{q_{2}} & 0 & 0 & 0 & 0 & \frac{q_{1}}{q_{2}} & 0 & 0 \\
0 & \frac{1}{q_{1}} & 0 & 0 & 0 & \frac{q_{2}}{q_{1}} & 0 & 0 & 0 \\
0 & 0 & \frac{q_{1}}{q_{2}} & \frac{1}{q_{2}} & 0 & 0 & 0 & 0 & 0 \\
\frac{1}{q_{2}^{2}} & 0 & 0 & 0 & \frac{1}{q_{1}^{2}} & 0 & 0 & 0 & 0 \\
\end{array}
\right) ,
\end{eqnarray}
where $q_{1}=e^{i\varphi_{1}}$ and $q_{2}=e^{i\varphi_{2}}$, with the parameters $%
\varphi_{1}$ and $\varphi_{2}$ both are real.

The Gell-mann matrices $\lambda_{u}$ satisfy $%
[I_{\lambda},I_{\mu}]=if_{\lambda\mu\nu}I_{\nu}
(\lambda,\mu,\nu=1,\cdot \cdot \cdot ,8)$, where
$I_{\mu}=\frac{1}{2}\lambda_{\mu}$. To the later
convenience, we denote $I_{\lambda}$ by, $I_{\pm}=I_{1}\pm iI_{2}$, $%
V_{\pm}=I_{4}\mp iI_{5}$,$U_{\pm}=I_{6}\pm iI_{7}$,
$Y=\frac{2}{\sqrt{3}}I_{8} $. Introducing three sets of $SU(3)$
realizations
\begin{eqnarray}\label{16}
(i):\left\{
\begin{array}{lll}
I_{\pm}^{(1)}=I_{1}^{\pm}I_{2}^{\pm},~~~U_{\pm}^{(1)}=U_{1}^{\pm}U_{2}^{
\pm},~~~V_{\pm}^{(1)}=V_{1}^{\pm}V_{2}^{\pm},\\
&\\
I_{3}^{(1)}=\frac{1}{3}(I_{1}^{3}+I_{2}^{3})+\frac{1}{2}%
(I_{1}^{3}Y_{2}+Y_{1}I_{2}^{3}),\\
&\\
Y^{(1)}=\frac{1}{3}(Y_{1}+Y_{2})+\frac{2}{3}I_{1}^{3}I_{2}^{3}-\frac{1}{2}
Y_{1}Y_{2};
\end{array}
\right.
\end{eqnarray}
\begin{eqnarray}\label{17}
(ii):\left\{
\begin{array}{lll}
I_{\pm}^{(2)}=U_{1}^{\pm}V_{2}^{\pm},~~~U_{\pm}^{(2)}=V_{1}^{\pm}I_{2}^{%
\pm},~~~V_{\pm}^{(2)}=I_{1}^{\pm}U_{2}^{\pm} , \\
& \\
I_{3}^{(2)}=\frac{1}{2}[-\frac{1}{3}(I_{1}^{3}+I_{2}^{3})+\frac{1}{2}%
(Y_{1}-Y_{2})+I_{1}^{3}Y_{2}+Y_{1}I_{2}^{3}], \\
&\\
Y^{(2)}=-\frac{1}{3}(I_{1}^{3}-I_{2}^{3})-\frac{1}{6}(Y_{1}+Y_{2})+\frac{2}{3%
}I_{1}^{3}I_{2}^{3}-\frac{1}{2}Y_{1}Y_{2} ;&\\
\end{array}
\right.
\end{eqnarray}
\begin{eqnarray}\label{18}
(iii):\left\{
\begin{array}{lll}
I_{\pm}^{(3)}=V_{1}^{\pm}U_{2}^{\pm},~~~U_{\pm}^{(3)}=I_{1}^{\pm}V_{2}^{%
\pm},~~~V_{\pm}^{(3)}=U_{1}^{\pm}I_{2}^{\pm}, \\
&\\
I_{3}^{(3)}=\frac{1}{2}[-\frac{1}{3}(I_{1}^{3}+I_{2}^{3})-\frac{1}{2}%
(Y_{1}-Y_{2})+I_{1}^{3}Y_{2}+Y_{1}I_{2}^{3}],\\
&\\
Y^{(3)}=\frac{1}{3}(I_{1}^{3}-I_{2}^{3})-\frac{1}{6}(Y_{1}+Y_{2})+\frac{2}{3}%
I_{1}^{3}I_{2}^{3}-\frac{1}{2}Y_{1}Y_{2},\\
\end{array}
\right.
\end{eqnarray}
where $I^{(k)}_{\pm}=I^{(k)}_{1}\pm iI^{(k)}_{2}$, $V^{(k)}_{%
\pm}=I^{(k)}_{4}\mp iI^{(k)}_{5}$,$U^{(k)}_{\pm}=I^{(k)}_{6}\pm
iI^{(k)}_{7}$, $Y^{(k)}=\frac{2}{\sqrt{3}}I^{(k)}_{8}$$(k=1,2,3)$,
which all satisfy the commutation relation $[I^{(i)}_{\lambda},I^{(j)}_{\mu}]=i\delta_{ij}f_{%
\lambda\mu\nu}I^{(i)}_{\nu} (\lambda,\mu,\nu=1,\cdot \cdot \cdot
,8;i,j=1,2,3)$.

For $i$-th and $(i+1)$-th lattices, in terms of the above operators
(\ref{16}) (\ref{17}) (\ref{18}), $M$ can be expressed as follows,
\begin{eqnarray}
M&=&\frac{q_{2}^{2}}{q_{1}^{2}}I_{+}^{(1)}+\frac{q_{1}^{2}}{q_{2}^{2}}I_{-}^{(1)}+\frac{1}{q_{2}^{2}}V_{+}^{(1)}%
+q_{2}^{2}V_{-}^{(1)}+q_{1}^{2}U_{+}^{(1)}+\frac{1}{q_{1}^{2}}U_{-}^{(1)}  \nonumber\\
&+&\frac{q_{1}}{q_{2}}I_{+}^{(2)}+\frac{q_{2}}{q_{1}}I_{-}^{(2)}+q_{2}V_{+}^{(2)}%
+\frac{1}{q_{2}}V_{-}^{(2)}+\frac{1}{q_{1}}U_{+}^{(2)}+q_{1}U_{-}^{(2)}  \nonumber\\
&+&\frac{q_{1}}{q_{2}}I_{+}^{(3)}+\frac{q_{2}}{q_{1}}I_{-}^{(3)}+q_{2}V_{+}^{(3)}%
+\frac{1}{q_{2}}V_{-}^{(3)}+\frac{1}{q_{1}}U_{+}^{(3)}+q_{1}U_{-}^{(3)}.
\end{eqnarray}
We eventually arrive at the unitary Yang-Baxter matrix for two
qutrits as,

\begin{equation}
\breve{R}_{i}(x)=\frac{1}{3}(b\mathbf{1}_{i}+aM_{i}),  \label{6}
\end{equation}
where $a=x^{-1}-x$ and $b=2x+x^{-1}$. So the whole tensor space
$\mathbb{C}^{3}\otimes\mathbb{C}^{3}$ is completely decomposed into
three subspaces. i.e. $
\mathbb{C}^{3}\otimes\mathbb{C}^{3}=\mathbb{C}^{3}\oplus
\mathbb{C}^{3}\oplus \mathbb{C}^{3}$. In addition, each block of
$\breve{R}$-matrix can be represented by fundamental representations
of SU(3) algebra.

When one acts $\breve{R}_{i}(x)$ directly on the separable state $|00\rangle$%
, he yields the following family of states
\begin{equation}
|\psi\rangle_{YB}=\frac{1}{3}(b|00\rangle+\frac{aq_{1}^{2}}{q_{2}^{2}}|11\rangle+%
\frac{a}{q_{2}^{2}}|22\rangle).
\end{equation}
By means of negativity, we study these entangled states. The
negativity for two qutrits is given by,
\begin{equation}
    \mathscr{N}(\rho)\equiv\frac{\|\rho^{T_{A}}\|-1}{2},
\end{equation}
where $\|\rho^{T_{A}}\|$ denotes the trace norm of $\rho^{T_{A}}$,
which denotes the partial transpose of the bipartite state $\rho$
\cite{Zyczkowski}. In fact, $\mathscr{N}(\rho)$ corresponds to the
absolute value of the sum of negative eigenvalues of $\rho^{T_{A}}$,
and negativity vanishes for unentangled states \cite{G. Vidal}. By
calculation, we can obtain the negativity of the state
$|\psi\rangle_{YB}$ as
\begin{equation}\label{N}
\mathscr{N}(\theta)=\frac{4}{9}(\sin^{2}\theta+|\sin\theta|\sqrt{%
1+8\cos^{2}\theta}).
\end{equation}
When $x=e^{i\frac{\pi}{3}}$, the state $|\psi\rangle_{YB}$ becomes
the maximally entangled state of two
qutrits as $|\psi\rangle_{YB}=\frac{-i}{\sqrt{3}}%
(e^{i\frac{2\pi}{3}}|00\rangle+\frac{q_{1}^{2}}{q_{2}^{2}}|11\rangle+\frac{1}{q_{2}^{2}}|22\rangle)$.
In general, if one acts the unitary $\breve{R}(x)$ on the
basis ${|00\rangle,|01\rangle,|02\rangle,|10\rangle,|11\rangle,|12%
\rangle,|20\rangle,|21\rangle,|22\rangle}$, he will obtain the same
negativity as Eq. (\ref{N}). It is easy to check that the negativity
ranges from 0 to 1 when the parameter $\theta$ runs from 0 to $\pi$.
But for $\theta \in [0,\pi]$, the entanglement is not a monotonic
function of $\theta$. And when $x=e^{i\frac{\pi}{3}}$, one will
generate nine complete and orthogonal maximally entangled states of
two qutrits.

It is worth mention that the entanglement doesn't depend on the
parameters $\varphi_{1}$ and $\varphi_{2}$. So one can verify that
the parameters $\varphi_{1}$ and $\varphi_{2}$ may be absorbed into
a local operation. Actually, we can introduce a local unitary
operation $P$ whose form is $P=diag\{\frac{q_{1}}{q_{2}},1,q_{1}\}$.
By means of this local transformation $(P\otimes
P)\breve{R}(\theta,\varphi_{1},\varphi_{2})(P^{-1}\otimes
P^{-1})=\breve{R}(\theta)$, we can say the unitary
$\breve{R}(\theta,\varphi_{1},\varphi_{2})$-matrix (\ref{6}) is
local equivalent to the universal  $\breve{R}(\theta)$-matrix, which
is the solution of $n=3$ in Ref.\cite{cxg1}, where the proof of
universality for a $n^{2}\times n^{2}$ Yang-Baxter matrix has been
presented. So the same as the property of $\breve{R}(\theta)$-matrix
in Ref.\cite{cxg1}, we can also say all pure entangled states of two
3-dimensional quantum systems (\emph{i.e.}, two qutrits) can be
generated from an initial separable state via the universal
$\breve{R}(\theta,\varphi_{1},\varphi_{2})$-matrix (\ref{6}) if one
is assisted by local unitary transformations.

\section{Hamiltonian and Berry Phase}\label{sec3}

\label{sec3} A Hamiltonian of the Yang-Baxter system can be
constructed from the $\breve{R}(\theta ,
\varphi_{1},\varphi_{2})$-matrix. As is shown in Ref. \cite{ckg},
the Hamiltonian is obtained through the Schr\"{o}dinger evolution of
the entangled states. Let the parameters $\theta$ be
time-independent and $\varphi_{i}=\omega_{i} t$ be time-dependent,
where $\omega_{i}=n_{i}\omega$(i=1,2;$\frac{n_{1}}{n_{2}}$ is a
fraction in lowest terms), the Hamiltonian reads,
\begin{eqnarray}
\hat{H} &=& i\hbar\frac{\partial\breve{R}(\theta ,
\varphi_{1},\varphi_{2})}{\partial t}\breve{R}^{\dag }(\theta ,
\varphi_{1},\varphi_{2}) =\bigoplus _{k=1}^{3}H^{(k)},
\end{eqnarray}
where the superscript $k$ denotes the $k$-th subsystem. The $k$-th
subsystem's Hamiltonian $H^{(k)}$ can be expressed as follows,
\begin{eqnarray}\label{H1}
  H^{(1)}=-\frac{8\sqrt{2}\hbar \omega \sin\theta}{3}&&[\frac{\sqrt{2}}{12}%
(ib^{*}n_{1}-ibn_{2})e^{-2i(\varphi_{1}-\varphi_{2})}I_{+}^{(1)}+\frac{\sqrt{%
2}}{12}(-ibn_{1}+ib^{*}n_{2})e^{2i(\varphi_{1}-\varphi_{2})}I_{-}^{(1)} \nonumber\\
&&+\frac{\sqrt{2}}{12}(ib^{*}n_{2}-2n_{1}\sin\theta)e^{-2i%
\varphi_{2}}V_{+}^{(1)}-\frac{\sqrt{2}}{12}(ibn_{2}+2n_{1}\sin\theta)e^{2i%
\varphi_{2}}V_{-}^{(1)}  \nonumber\\
&&-\frac{\sqrt{2}}{12}(ibn_{1}+2n_{2}\sin\theta)e^{2i\varphi_{1}}U_{+}^{(1)}+%
\frac{\sqrt{2}}{12}(ib^{*}n_{1}-2n_{2}\sin\theta)e^{-2i%
\varphi_{1}}U_{-}^{(1)}  \nonumber\\
&&-\frac{\sqrt{2}}{2}(n_{1}-n_{2})\sin\theta I_{3}^{(1)}+\frac{\sqrt{2}}{4}%
(n_{1}+n_{2})\sin\theta Y^{(1)}]  ,
\end{eqnarray}
\begin{eqnarray}\label{H2}
 H^{(i)}=-\frac{4\sqrt{2}\hbar \omega \sin\theta}{3}&&[\frac{\sqrt{2}}{12}%
(ibn_{2}-ib^{*}n_{1})e^{i(\varphi_{1}-\varphi_{2})}I_{+}^{(i)}+\frac{\sqrt{2}%
}{12}(-ib^{*}n_{2}+ibn_{1})e^{-i(\varphi_{1}-\varphi_{2})}I_{-}^{(i)} \nonumber\\
&&+\frac{\sqrt{2}}{12}(2n_{1}\sin\theta-ib^{*}n_{2})e^{i%
\varphi_{2}}V_{+}^{(i)}+\frac{\sqrt{2}}{12}(2n_{1}\sin\theta+ib%
n_{2})e^{-i\varphi_{2}}V_{-}^{(i)}   \nonumber\\
&&+\frac{\sqrt{2}}{12}(ibn_{1}+2n_{2}\sin\theta)e^{-i\varphi_{1}}U_{+}^{(i)}+%
\frac{\sqrt{2}}{12}(-ib^{*}n_{1}+2n_{2}\sin\theta)e^{i\varphi_{1}}U_{-}^{(i)}
 \nonumber\\
&&+\frac{\sqrt{2}}{2}(n_{1}-n_{2})\sin\theta I_{3}^{(i)}-\frac{\sqrt{2}}{4}%
(n_{1}+n_{2})\sin\theta Y^{(i)}].
\end{eqnarray}
Hereafter the superscript $i$($i=2,3$) denotes the second or the
third subsystem.

 Based on the operators $I_{\lambda}^{(k)}$($\lambda=1,2,\cdots,8; k=1,2,3$),
The $k$-th subsystem's Hamiltonian can be rewritten as follows,
\begin{eqnarray}\label{H11}
H^{(k)}=C(k)&&\sum_{\lambda=1}^{8}B_{%
\lambda}^{(k)}I_{\lambda}^{(k)},
\end{eqnarray}
where $C(1)=-\frac{8\sqrt{2}\hbar \omega \sin\theta}{3}$ and
$C(i)=-\frac{4\sqrt{2}\hbar \omega \sin\theta}{3}$($i=2,3$). By
comparing Eq(\ref{H1}), Eq(\ref{H2}) with Eq(\ref{H11}), one can
obtain $B_{\lambda}^{(k)}$ as follows,
\begin{eqnarray}
\left\{
\begin{array}{llllllll}
B_{1}^{(1)} = \frac{\sqrt{2}}{2}(n_{1}-n_{2})\cos\theta
\sin2(\varphi_{1}-\varphi_{2})+\frac{\sqrt{2}}{6}(n_{1}+n_{2})\sin\theta
\cos2(\varphi_{1}-\varphi_{2}),\\
&\\
B_{2}^{(1)} = -\frac{\sqrt{2}}{2}(n_{1}-n_{2})\cos\theta
\cos2(\varphi_{1}-\varphi_{2})+\frac{\sqrt{2}}{6}(n_{1}+n_{2})\sin\theta
\sin2(\varphi_{1}-\varphi_{2}),\\
&\\
B_{3}^{(1)}=-\frac{\sqrt{2}}{2}(n_{1}-n_{2})\sin\theta,\\
&\\
B_{4}^{(1)} = \frac{\sqrt{2}}{6}n_{2}\sin\theta\cos(2\varphi_{2})+\frac{%
\sqrt{2}}{2}n_{2}\cos\theta \sin(2\varphi_{2}) -\frac{\sqrt{2}}{3}%
n_{1}\sin\theta\cos(2\varphi_{2}),\\
&\\
B_{5}^{(1)} = -\frac{\sqrt{2}}{6}n_{2}\sin\theta\sin(2\varphi_{2})+\frac{%
\sqrt{2}}{2}n_{2}\cos\theta\cos(2\varphi_{2}) +\frac{\sqrt{2}}{3}%
n_{1}\sin\theta \sin(2\varphi_{2}),\\
&\\
B_{6}^{(1)} = \frac{\sqrt{2}}{6}n_{1}\sin\theta\cos(2\varphi_{1})+\frac{%
\sqrt{2}}{2}n_{1}\cos\theta \sin(2\varphi_{1}) -\frac{\sqrt{2}}{3}%
n_{2}\sin\theta\cos(2\varphi_{1}),\\
&\\
B_{7}^{(1)} = -\frac{\sqrt{2}}{6}n_{1}\sin\theta\sin(2\varphi_{1})+\frac{%
\sqrt{2}}{2}n_{1}\cos\theta\cos(2\varphi_{1}) +\frac{\sqrt{2}}{3}%
n_{2}\sin\theta \sin(2\varphi_{1}),\\
&\\
B_{8}^{(1)}=\frac{\sqrt{6}}{6}(n_{1}+n_{2})\sin\theta,
\end{array}
\right.
\end{eqnarray}
\begin{eqnarray}
\left\{
\begin{array}{llllllll}
B_{1}^{(i)} = \frac{\sqrt{2}}{2}(n_{1}-n_{2})
\sin(\varphi_{1}-\varphi_{2})\cos\theta-\frac{\sqrt{2}}{6}(n_{1}+n_{2})\cos(%
\varphi_{1}-\varphi_{2})\sin\theta,\\
&\\
B_{2}^{(i)} = \frac{\sqrt{2}}{2}(n_{1}-n_{2})
\cos(\varphi_{1}-\varphi_{2})\cos\theta+\frac{\sqrt{2}}{6}(n_{1}+n_{2})
\sin(\varphi_{1}-\varphi_{2})\sin\theta,\\
&\\
B_{3}^{(i)}=\frac{\sqrt{2}}{2}(n_{1}-n_{2})\sin\theta,\\
&\\
B_{4}^{(i)} = -\frac{\sqrt{2}}{6}n_{2}
\cos\varphi_{2}\sin\theta+\frac{\sqrt{2}}{2}n_{2}
\sin\varphi_{2}\cos\theta +\frac{\sqrt{2}}{3}n_{1}
\cos\varphi_{2}\sin\theta,\\
&\\
B_{5}^{(i)} = -\frac{\sqrt{2}}{6}n_{2}\sin\varphi_{2}\sin\theta-\frac{%
\sqrt{2}}{2}n_{2} \cos\varphi_{2}\cos\theta +\frac{\sqrt{2}}{3}n_{1}
\sin\varphi_{2}\sin\theta,\\
&\\
B_{6}^{(i)} = -\frac{\sqrt{2}}{6}n_{1} \cos\varphi_{1}\sin\theta+\frac{%
\sqrt{2}}{2}n_{1} \sin\varphi_{1}\cos\theta +\frac{\sqrt{2}}{3}n_{2}
\cos\varphi_{1}\sin\theta,\\
&\\
B_{7}^{(i)} = -\frac{\sqrt{2}}{6}n_{1} \sin\varphi_{1}\sin\theta-\frac{%
\sqrt{2}}{2}n_{1} \cos\varphi_{1}\cos\theta +\frac{\sqrt{2}}{3}n_{2}
\sin\varphi_{1}\sin\theta,\\
&\\
B_{8}^{(i)}=-\frac{\sqrt{6}}{6}(n_{1}+n_{2})\sin\theta.
\end{array}
\right.
\end{eqnarray}
The Hamiltonian of the $k$-th subsystem,
$H(\textbf{B}(t)^{(k)})^{(k)}$, depends on the parameters
$B_{\lambda}^{(k)}$($\lambda=1,2,\cdots,8$), which are the
components of vector $\textbf{B}^{(k)}$. Namely, $\textbf{B}^{(k)}$
are a series of time-varying parameters controlling the $k$-th
subsystem's Hamiltonian. After a periods $T^{(k)}$, Hamiltonian
returns to its original form, i.e.
$H(\textbf{B}(0))^{(k)}=H(\textbf{B}(T^{(k)}))^{^{(k)}}$. According
to this, one can easily verify the periods of the subsystems are,
$T^{(1)}=\frac{\pi}{\omega}$ and $T^{(i)}=\frac{2\pi}{\omega}$ when
$\frac{n_{1}}{n_{2}}=m$(i.e. $n_{1}=m, n_{2}=1$, where $m$ is
integer and $m\neq0$).

  To the later convenience, we denote $\sqrt{n_{1}^{2}-%
n_{1}n_{2}+n_{2}^{2}}= n$, $%
3n_{1}+2\sqrt{2}n\sin\theta=\alpha_{+}$, $3n_{1}+%
\sqrt{2}ib^{*}n=\beta_{+}$, $3n_{1}-2\sqrt{2}%
n\sin\theta=\alpha_{-}$, $3n_{1}-\sqrt{2}%
ib^{*}n=\beta_{-}$, and $3n_{2}+2\sqrt{2}%
n\sin\theta=\delta_{+}$, $3n_{2}+\sqrt{2}%
ibn=\eta_{+}$, $3n_{2}-2\sqrt{2}n\sin\theta=%
\delta_{-}$, $3n_{2}-\sqrt{2}ibn=\eta_{-}$.
$\mathcal{N}_{\alpha}^{(k)}$'s$(\alpha=+,0,-)$ are normalization
coefficients. $
\mathcal{N}^{(1)}_{+}=12n^{2}[6n^{2}+2\sqrt{2}(n_{1}-2n_{2})n
\sin\theta-3n_{1}^{2}]$, $\mathcal{N}^{(1)}_{0}=2n^{2}$,
$\mathcal{N} ^{(1)}_{-}=12n^{2}[6n^{2}-2\sqrt{2}(n_{1}-2n_{2})n
\sin\theta-3n_{1}^{2}]$; $
\mathcal{N}^{(i)}_{+}=12n^{2}[3n^{2}-2\sqrt{2}(n_{1}+n_{2})n
\sin\theta+3n_{1}n_{2}]$, $\mathcal{N}^{(i)}_{0}=2n^{2}$,
$\mathcal{N} ^{(i)}_{-}=12n^{2}[3n^{2}+2\sqrt{2}(n_{1}+n_{2})n
\sin\theta+3n_{1}n_{2}]$. The eigenstates of the first subsystem are
found to be
\begin{eqnarray}\label{8}
 \left\{
\begin{array}{lll}
|E^{(1)}_{+}\rangle &=& \frac{1}{\sqrt{\mathcal{N}_{+}^{(1)}}}%
(((n_{1}-2n_{2})\alpha_{+}+6n_{2}^{2})e^{2i\varphi_{2}}|00\rangle+(n_{2}%
\beta_{-}+\sqrt{2}ibnn_{1})e^{2i\varphi_{1}}|11\rangle+(n_{1}%
\alpha_{+}-n_{2}\beta_{+})|22\rangle),  \\
&\\
 |E^{(1)}_{0}\rangle &=& \frac{1}{\sqrt{\mathcal{N}_{0}^{(1)}}}%
(-n_{1}e^{2i\varphi_{2}}|00\rangle+n_{2}e^{2i\varphi_{1}}|11%
\rangle+(n_{1}-n_{2})|22\rangle) , \\
&\\
|E^{(1)}_{-}\rangle &=& \frac{1}{\sqrt{\mathcal{N}_{-}^{(1)}}}%
(((n_{1}-2n_{2})\alpha_{-}+6n_{2}^{2})e^{2i\varphi_{2}}|00\rangle+(n_{2}%
\beta_{+}-\sqrt{2}ibnn_{1})e^{2i\varphi_{1}}|11\rangle+(n_{1}%
\alpha_{-}-n_{2}\beta_{-})|22\rangle),
\end{array}
\right.
\end{eqnarray}
with the corresponding eigenvalues $E^{(1)}_{+} = \frac{4\sqrt{2}}{3}%
\hbar n\omega\sin\theta$, $E^{(1)}_{0} = 0$ and $E^{(1)}_{-} = -\frac{4\sqrt{2}%
}{3}\hbar n\omega\sin\theta$. For the second and the third
subsystems, the eigenstates are found to be,
\begin{eqnarray}\label{11}
 \left\{
\begin{array}{lll}
|E^{(2)}_{+}\rangle &=& \frac{1}{\sqrt{\mathcal{N}_{+}^{(2)}}}%
((n_{1}\alpha_{-}+n_{2}\delta_{-})e^{i\varphi_{2}}|01\rangle+(n_{1}%
\alpha_{-}-n_{2}\beta_{-}^{*})|12\rangle+(n_{2}%
\delta_{-}-n_{1}\eta_{+})e^{-i(\varphi_{1}-\varphi_{2})}|20\rangle) ,\\
&\\
|E^{(2)}_{0}\rangle &=& \frac{1}{\sqrt{\mathcal{N}_{0}^{(2)}}}%
((-n_{1}+n_{2})e^{i\varphi_{2}}|01\rangle+n_{1}|12\rangle-n_{2}e^{-i(\varphi_{1}-%
\varphi_{2})}|20\rangle), \\
&\\
|E^{(2)}_{-}\rangle &=& \frac{1}{\sqrt{\mathcal{N}_{-}^{(2)}}}%
((n_{1}\alpha_{+}+n_{2}\delta_{+})e^{i\varphi_{2}}|01\rangle+(n_{1}%
\alpha_{+}-n_{2}\beta_{+}^{*})|12\rangle+(n_{2}%
\delta_{+}-n_{1}\eta_{-})e^{-i(\varphi_{1}-\varphi_{2})}|20\rangle)
,
\end{array}
\right.
\end{eqnarray}
\begin{eqnarray}
 \left\{
\begin{array}{lll}\label{13}
|E^{(3)}_{+}\rangle &=& \frac{1}{\sqrt{\mathcal{N}_{+}^{(3)}}}%
((n_{2}%
\delta_{-}-n_{1}\eta_{+})e^{-i(\varphi_{1}-\varphi_{2})}|02\rangle+(n_{1}\alpha_{-}+n_{2}\delta_{-})e^{i\varphi_{2}}|10\rangle+(n_{1}%
\alpha_{-}-n_{2}\beta_{-}^{*})|21\rangle) ,\\
&\\
|E^{(3)}_{0}\rangle &=& \frac{1}{\sqrt{\mathcal{N}_{0}^{(3)}}}%
(-n_{2}e^{-i(\varphi_{1}-%
\varphi_{2})}|02\rangle+(-n_{1}+n_{2})e^{i\varphi_{2}}|10\rangle+n_{1}|21\rangle) ,\\
&\\
|E^{(3)}_{-}\rangle &=& \frac{1}{\sqrt{\mathcal{N}_{-}^{(3)}}}%
((n_{2}%
\delta_{+}-n_{1}\eta_{-})e^{-i(\varphi_{1}-\varphi_{2})}|02\rangle+(n_{1}\alpha_{+}+n_{2}\delta_{+})e^{i\varphi_{2}}|10\rangle+(n_{1}%
\alpha_{+}-n_{2}\beta_{+}^{*})|21\rangle),
\end{array}
\right.
\end{eqnarray}
 with the corresponding eigenvalues
$E^{(i)}_{+} = \frac{2\sqrt{2}}{3} \hbar n\omega\sin\theta$,
$E^{(i)}_{0} = 0$ and $E^{(i)}_{-} = -\frac{2\sqrt{2} }{3}\hbar
n\omega\sin\theta$ ($i=2,3$).

 According to the definition of Berry
phase \cite{berry}, when the parameter $\textbf{B}^{(k)}$ is slowly
changed around a circuit, then at the end of circuit, the
eigenstates $|E_{\alpha}^{(k)}\rangle$($\alpha=+,0,-$ )evolves
adiabatically from 0 to $T^{(k)}$, the Berry phases accumulated by
the states $|E_{\alpha}^{(k)}\rangle$ are,

\begin{equation}
\gamma _{\alpha}^{(k)}=i\int_{0}^{T^{(k)}}\langle E_{\alpha }^{(k)}|
\frac{\partial}{\partial t}|E_{\alpha }^{(k)}\rangle dt.
\label{berryphase}
\end{equation}%
By substituting (\ref{8}) (\ref{11}) (\ref{13}) into
(\ref{berryphase}), one can obtain the Berry phases for these
eigenstates,
\begin{equation}\label{7}
 Subsystem~1:\left\{
\begin{array}{lll}
\mathcal{\gamma
}_{+}^{(1)}=\frac{K_{1}+K_{2}}{\mathcal{N}_{+}^{(1)}}2\omega T^{(1)}\\
& \\
\mathcal{\gamma
}_{0}^{(1)}=-\frac{n_{1}n_{2}(n_{1}+n_{2})}{2n^{2}}2\omega T^{(1)}\\
&\\
\mathcal{\gamma
}_{-}^{(1)}=\frac{K_{1}-K_{2}}{\mathcal{N}_{-}^{(2)}}2\omega T^{(1)}
\end{array}
\right. ~~Subsystem~2~or~3:\left\{
\begin{array}{lll}
\mathcal{\gamma}_{+}^{(i)}=\frac{K_{3}+K_{4}}{\mathcal{N}_{+}^{(i)}}\omega T^{(i)}\\
& \\
\mathcal{\gamma}_{0}^{(i)}=-\frac{n_{2}(n_{1}-n_{2})(n_{1}-2n_{2})}{2n^{2}}\omega T^{(i)}\\
&\\
\mathcal{\gamma}_{-}^{(i)}=\frac{K_{3}-K_{4}}{\mathcal{N}_{-}^{(i)}}\omega
T^{(i)}
\end{array}
\right.
\end{equation}
where
\begin{eqnarray*}
K_{1}
&=&-10n_{1}^{5}+13n_{1}^{4}n_{2}+11n_{1}^{3}n_{2}^{2}-82n_{1}^{2}n_{2}^{3}+94n_{1}n_{2}^{4}-52n_{2}^{5}-8\cos
(2\theta )(n_{1}-2n_{2})n^{4},
\\
K_{2} &=&-6\sqrt{2}\sin \theta
nn_{2}(n_{1}^{3}-9n_{1}^{2}n_{2}+12n_{1}n_{2}^{2}-8n_{2}^{3}) ,\\
K_{3} &=&2n^{2}(9n_{1}^{2}(n_{1}-n_{2})-8\sin^{2}\theta
(n_{1}^{3}+n_{2}^{3}))-9n_{2}(n_{1}^{4}-n_{1}^{3}n_{2}+5n_{1}^{2}n_{2}^{2}-3n_{1}n_{2}^{3}+2n_{2}^{4}),
\\
K_{4} &=&6\sqrt{2}\sin \theta
nn_{2}(n_{1}^{3}+6n_{1}^{2}n_{2}-3n_{1}n_{2}^{2}+4n_{2}^{3}).
\end{eqnarray*}
The above Berry phases are the general result for the whole system.
Despite the complexity in the expressions for the Berry phase in its
full generality, we can use the general formulae to discuss some
special cases for convenience to understand.
\\
\\
$\mathbf {Example ~1}$: For $\varphi_{1}=\varphi_{2}$, i.e.
$n_{1}=n_{2}=1$. In this case, $T^{(1)}=\frac{ \pi}{\omega}$ and
$T^{(i)}=\frac{2\pi}{\omega}$.
 When one substitutes the conditions
into Eq. (\ref{7}), he then gets the following explicit expressions
of the Berry phase(all phases are defined modulo 2$\pi$ throughout
this paper),
\begin{equation}\label{9}
\left\{
\begin{array}{ll}
\mathcal{\gamma }_{+}^{(k)}=-\mathcal{\gamma
}_{-}^{(k)}=-\pi(1-\frac{2\sqrt{2}}{3}\sin
\theta),\\
&\\
 \mathcal{\gamma }_{0}^{(k)}=0,
\end{array}
\right.
\end{equation}
where $k=1,2,3$. In fact, for $ \varphi_{1}=\varphi_{2}$, the
Hamiltonian can be represented based on three sets of $SU(2)$
operators. Under this framework, the Berry phase can be interpreted.
In the following, we have a thorough discussing for the case.

  Introducing three sets
of SU(2) realizations in terms of three sets of operators (\ref{16})
(\ref{17}) (\ref{18}),
\begin{eqnarray}\label{rlz1}
\left\{
\begin{array}{lll}
S_{+}^{(k)} & = & \frac{1}{\sqrt{2}}(V_{-}^{(k)}+U_{+}^{(k)})\\
&\\
S_{-}^{(k)} & = & \frac{1}{\sqrt{2}}(V_{+}^{(k)}+U_{-}^{(k)})\\
&\\
S_{3}^{(k)} & = &
\frac{3}{4}Y^{(k)}+\frac{1}{4}(I_{+}^{(k)}+I_{-}^{(k)}).
\end{array}
\right.
\end{eqnarray}
They satisfy the algebra relations of $SU(2)$ group:
$[S^{(i)}_{+},S^{(j)}_{-}]=2 \delta_{ij}S^{(i)}_{3}$,
$[S^{(i)}_{3},S^{(j)}_{\pm}]=\pm \delta_{ij}S^{(i)}_{\pm}$,
$(S_{\pm}^{(i)})^{2}=0$($i,j=1,2,3$), with
$S^{(k)}_{\pm}=S^{(k)}_{1}\pm iS^{(k)}_{2}$($k=1,2,3$). By the way,
their second-order Casimir operators are $\mathcal
{J}^{(k)}=\frac{1}{2}(S^{(k)}_{+}S^{(k)}_{-}+S^{(k)}_{-}S^{(k)}_{+})+(S^{(k)}_{3})^{2}$.
One can verify that the eigenvalues of  $\mathcal {J}^{(k)}$ are
$\frac{1}{2}(\frac{1}{2}+1)=\frac{3}{4}$ and $0(0+1)=0$ which
correspond to spin-$\frac{1}{2}$ system and spin-0 system. In terms
of the operators (\ref{rlz1}), the Hamiltonian for subsystems Eq.
(\ref{H1}) and Eq. (\ref{H2}) can be rewritten as follows:
\begin{eqnarray}\label{eq4.1}
 H^{(1)}=C(1)(-\frac{1}{6}ib^{*}e^{2i\varphi}
S_{+}^{(1)}+\frac{1}{6}ibe^{-2i\varphi}
S_{-}^{(1)}+\frac{2\sqrt{2}}{3}\sin\theta S_{3}^{(1)}),
\end{eqnarray}
\begin{eqnarray}\label{eq4.2}
 H^{(2)}=C(2)(-\frac{1}{6}ibe^{i\varphi}
S_{+}^{(2)}+\frac{1}{6}ib^{*}e^{-i\varphi}
S_{-}^{(2)}+\frac{2\sqrt{2}}{3}\sin\theta S_{3}^{(2)}),
\end{eqnarray}
\begin{eqnarray}\label{eq4.3}
 H^{(3)}=C(3)(-\frac{1}{6}ibe^{i\varphi}
S_{+}^{(3)}+\frac{1}{6}ib^{*}e^{-i\varphi}
S_{-}^{(3)}+\frac{2\sqrt{2}}{3}\sin\theta S_{3}^{(3)}).
\end{eqnarray}
$C(1)$, $C(2)$ and $C(3)$ are defined in Eq. (\ref{H11}). So we can
say the whole system is equivalent to three spin-$\frac{1}{2}$
subsystems and three spin-0 subsystems. Actually, we can introduce a
$9\times9$ orthogonal matrix P which is time-independent(see
Appendix A). By means of P, the whole system's Hamiltonian $\hat{H}$
and Casimir operators $\mathcal {J}^{(k)}$ are transformed into
block-diagonal matrices. Namely, $\tilde{\hat{H}}=P\hat{H}P^{T}$ and
$\tilde{\mathcal {J}}^{(k)}=P\mathcal {J}^{(k)}P^{T}$ are
block-diagonal matrices, where $P^{T}$ denotes the transpose of
matrix $P$.

For the subsystem 1, from  Eq. (\ref{12}) we can obtain its
Hamiltonian $\tilde{H}^{(1)}=H_{\frac{1}{2}}^{(1)}\oplus
H_{0}^{(1)}$. For $H_{0}^{(1)}$, the eigenvalue of Casimir operator
$\mathcal {J}^{(1)}$ is 0, and the Berry Phase is 0. So we can say
the subsystem Hamiltonian $H_{0}^{(1)}$ is equivalent to a spin-$0$
subsystem. For $H_{\frac{1}{2}}^{(1)}$, we introduce the
transformation, $\cos\alpha=\frac{2\sqrt{2}}{3}\sin\theta$ and
$\cos\beta=\frac{-\sin\theta \cos2\varphi+3\cos\theta
\sin2\varphi}{\sqrt{9-8\sin^{2}\theta}}$, where $\alpha \in
(\arccos\frac{2\sqrt{2}}{3},\arccos-\frac{2\sqrt{2}}{3})$ and $\beta
\in[0,2\pi]$. $\alpha$ is time-independent, and $\beta$ is
time-dependent. Then the Berry phase Eq(\ref{9}) can be recast as,
\begin{eqnarray}
\mathcal{\gamma}_{\pm}^{(1)}=\mp\pi(1-\cos\alpha)=\mp\frac{\Omega(C)}{2}
\end{eqnarray}
where $\Omega(C) = 2\pi (1- \cos\alpha)$ is the familiar solid angle
enclosed by the loop on the Bloch sphere, and the parameter $\alpha$
comes from $\theta$ which comes from the Yang-Baxterization of the
Hermitian matrix $M$. So the Berry phase depends on the spectral
parameter. By means of $P$, the eigenstates $|E^{(1)}_{\pm}\rangle$
can be recast as follows(we neglected the global phase factor),
\begin{eqnarray}
|E^{(1)}_{+}\rangle &=&
-e^{i\beta}\sin\frac{\alpha}{2}|00\rangle+\cos\frac{\alpha}{2}|01\rangle,\\
|E^{(1)}_{-}\rangle &=&
\cos\frac{\alpha}{2}|00\rangle+e^{-i\beta}\sin\frac{\alpha}{2}|01\rangle.
\end{eqnarray}
 The Hamiltonian
$H^{(1)}$ in Eq(\ref{eq4.1}) can be recast based on the operators
Eq(\ref{10}) to the form
\begin{eqnarray}\label{22}
H_{\frac{1}{2}}^{(1)}&=&-2\hbar\omega\cos\alpha(2\cos\alpha
\tilde{S}^{(1)}_{3}+\sin\alpha
e^{i\beta}\tilde{S}^{(1)}_{+}+\sin\alpha e^{-i\beta}\tilde{S}^{(1)}_{-})\nonumber\\
&=&-2\hbar\omega\cos\alpha \hat{H}_{0}^{(1)},
\end{eqnarray}
where $\hat{H}_{0}^{(1)}$ is of the following form,
\begin{eqnarray}
\hat{H}_{0}^{(1)}=2\cos\alpha \tilde{S}^{(1)}_{3}+\sin\alpha
e^{i\beta}\tilde{S}^{(1)}_{+}+\sin\alpha
e^{-i\beta}\tilde{S}^{(1)}_{-}.\nonumber
\end{eqnarray}
Thus, the Hamiltonian $H_{\frac{1}{2}}^{(1)}$ has the same physical
meaning as that given in \cite{ckg2}. That is,
$H_{\frac{1}{2}}^{(1)}$ is an oscillator Hamiltonian formed by two
fermions with frequency $2\omega\cos\alpha$ where $\alpha \in
(\arccos\frac{2\sqrt{2}}{3},\arccos-\frac{2\sqrt{2}}{3})$. For the
Yang-Baxter subsystem, $\alpha=0$ is a critical point. However,
because of $\alpha\neq 0$, the subsystem Hamiltonian Eq(\ref{22})
can't reduce to the standard oscillator. In other words, $\sin\alpha
e^{i\beta}$ plays a role of the "energy gap" and the wave function
of the subsystem takes the form of spin-coherent state\cite{smv}.
The quantum criticality can't occur in the Hamiltonian subsystem
$H_{\frac{1}{2}}^{(1)}$ and the Berry phases cant't vanish since
$\alpha\neq 0$.

 Via the same method, the Berry phases for
subsystems 2 and 3 may be obtained,
$\gamma_{\pm}^{(i)}=\mp\pi(1-\cos\alpha)=\mp\frac{\Omega(C)}{2}$ and
$\gamma_{0}^{(i)}=0$($i=2,3$). The Hamiltonian $H^{(i)}$ in
Eq(\ref{eq4.2}) and Eq(\ref{eq4.3}) can be recast to the form
\begin{eqnarray}
H_{\frac{1}{2}}^{(2)}=H_{\frac{1}{2}}^{(3)}=-\hbar\omega\cos\alpha\hat{H}_{0}^{(i)},
\end{eqnarray}
where $\hat{H}_{0}^{(i)}=2\cos\alpha \tilde{S}^{(i)}_{3}+\sin\alpha
e^{i\beta}\tilde{S}^{(i)}_{+}+\sin\alpha
e^{-i\beta}\tilde{S}^{(i)}_{-}$($i=2,3$). Thus, the Hamiltonian
$H_{\frac{1}{2}}^{(2)}$ and $H_{\frac{1}{2}}^{(3)}$ both have the
same physical meaning as that given in \cite{ckg2}. That is,
$H_{\frac{1}{2}}^{(2)}$ and $H_{\frac{1}{2}}^{(3)}$ both are
oscillator Hamiltonians formed by two fermions with frequency
$\omega\cos\alpha$. Thus, in the two Yang-Baxter subsystem
$H_{\frac{1}{2}}^{(2)}$ and $H_{\frac{1}{2}}^{(3)}$, the quantum
criticality can't occur and the Berry phases cant't vanish since
$\alpha\neq 0$.

$\mathbf{Example ~2}$: For $\varphi_{1}=-\varphi_{2}$, i.e.
$n_{1}=-n_{2}=-1$. In this case $T^{(1)}= \frac{\pi}{\omega}$ and
$T^{(i)}=\frac{2\pi}{\omega}$. One can get the following explicit
expressions of the Berry phase,
\begin{equation}
  Subsystem~1:\left\{
\begin{array}{lll}
\gamma^{(1)}_{+}=\frac{\sqrt{6}\sin\theta}{3}2\pi \\
& \\
\gamma^{(1)}_{0}=0\\
&\\
\gamma^{(1)}_{-}=-\frac{\sqrt{6}\sin\theta}{3}2\pi
\end{array}
\right.  ~~~~~~~~~Subsystem~2~or~3:\left\{
\begin{array}{lll}
\gamma^{(i)}_{+}=\frac{\sqrt{6}\sin\theta}{3}2\pi\\
& \\
\gamma^{(i)}_{0}= 0\\
&\\
\gamma^{(i)}_{-}=-\frac{\sqrt{6}\sin\theta}{3}2\pi
\end{array}
\right.
\end{equation}
In this case, one can find that
$\gamma^{(1)}_{+}=-\gamma^{(i)}_{-}$, $
\gamma^{(1)}_{0}=-\gamma^{(i)}_{0}$ and
$\gamma^{(1)}_{-}=-\gamma^{(i)}_{+}$.

$\mathbf{Example  ~3}$: For $\varphi_{1}=2\varphi_{2}$, i.e.
$n_{1}=2n_{2}=2$. In this case $T^{(1)}= \frac{\pi}{\omega}$ and
$T^{(i)}=\frac{2\pi}{\omega}$. We then gets the following explicit
expressions of the Berry phase,
\begin{equation}
 Subsystem~1:\left\{
\begin{array}{lll}
\gamma^{(1)}_{+}=\frac{\sqrt{6}\sin\theta}{3}2\pi \\
& \\
\gamma^{(1)}_{0}=0\\
&\\
\gamma^{(1)}_{-}=-\frac{\sqrt{6}\sin\theta}{3}2\pi
\end{array}
\right. ~~~~~~~~~Subsystem~2~or~3:\left\{
\begin{array}{lll}
\gamma^{(i)}_{+}=\frac{\sqrt{6}\sin\theta}{3}2\pi\\
& \\
\gamma^{(i)}_{0}= 0\\
&\\
\gamma^{(i)}_{-}=-\frac{\sqrt{6}\sin\theta}{3}2\pi
\end{array}
\right.
\end{equation}
In this case, one can find that
$\gamma^{(1)}_{+}=-\gamma^{(i)}_{-}$, $
\gamma^{(1)}_{0}=-\gamma^{(i)}_{0}$ and
$\gamma^{(1)}_{-}=-\gamma^{(i)}_{+}$.

$\mathbf {Example ~4}$: For $\varphi_{1}=-2\varphi_{2}$, i.e.
$n_{1}=-2n_{2}=-2$. In this case $T^{(1)}= \frac{\pi}{\omega}$ and
$T^{(i)}=\frac{2\pi}{\omega}$. We then get the following explicit
expressions of the Berry phase,
\begin{equation}
  Subsystem~1:\left\{
\begin{array}{lll}
\gamma^{(1)}_{+}&=&(\frac{4}{7}+\frac{\sqrt{14}}{3}\sin\theta)2\pi\\
& \\
 \gamma^{(1)}_{0}&=&-\frac{2\pi}{7}\\ & \\
\gamma^{(1)}_{-}&=&(\frac{4}{7}-\frac{\sqrt{14}}{3}\sin\theta)2\pi
\end{array}
\right. ~~~~~Subsystem~2~or~3:\left\{
\begin{array}{lll}
\gamma^{(i)}_{+}&=&-(\frac{4}{7}-\frac{\sqrt{14}}{3}\sin\theta)2\pi
  \\ & \\
 \gamma^{(i)}_{0}&=&\frac{2\pi}{7}\\ & \\
\gamma^{(i)}_{-}&=&-(\frac{4}{7}+\frac{\sqrt{14}}{3}\sin\theta)2\pi
\end{array}
\right.
\end{equation}
In this case, one can find that
$\gamma^{(1)}_{+}=-\gamma^{(i)}_{-}$, $
\gamma^{(1)}_{0}=-\gamma^{(i)}_{0}$ and
$\gamma^{(1)}_{-}=-\gamma^{(i)}_{+}$.

\section{Summary}\label{sec5}
In this paper, we have presented a $9 \times 9$ $M$-matrix which
satisfies the Hecke algebraic relations and derived a unitary
$\breve{R}(\theta,\varphi_{1},\varphi_{2})$-matrix via
Yang-Baxterization of the $M$-matrix. In the following, we can say
that all pure two-qutrit entangled states can be generated via the
universal $\breve{R}$-matrix assisted by local unitary
transformations. Specifically, the arbitrary degree of entanglement
for two-qutrit entangled states can be generated via the unitary
$\breve{R}$ matrix acting on the standard basis. Then the evolution
of the Yang-Baxter system is explored by constructing a Hamiltonian
from the unitary $\breve{R}$-matrix. In addition, the Berry phase of
the system is investigated, and general expressions of Berry phase
are figured out. Finally, we use the general expressions to discuss
some special cases. For $ \varphi_{1}=\varphi_{2}$, based on three
sets of SU(2) operators, the Hamiltonian has been represented and
the three nonzero Hamiltonian subsystems all are shown to be
equivalent to oscillator systems of two fermions. Under this
framework, the Berry phase can be interpreted.

\hspace{1cm}

 This work was supported in part by NSF of China (Grant
 No. 10875026).

\hspace{1cm}

\appendix

\section{Block-diagonalize $\hat{H}$ and $\mathcal {J}^{(k)}$ } \label{appendixa}
The $9\times9$ orthogonal matrix P which is time-independent reads,
\begin{eqnarray}
P=\left(
  \begin{array}{ccccccccc}
    \frac{1}{\sqrt{2}} & 0 & 0 & 0 & \frac{1}{\sqrt{2}}& 0 & 0 & 0 & 0 \\
     0 & 0 & 0 & 0 & 0 & 0 & 0 & 0 & 1 \\
    \frac{1}{\sqrt{2}} & 0 & 0 & 0 & -\frac{1}{\sqrt{2}} & 0 & 0 & 0 & 0 \\
     0 & 1 & 0 & 0 & 0 & 0 & 0 & 0 & 0 \\
     0 & 0 &0 & 0 & 0 & \frac{1}{\sqrt{2}} & \frac{1}{\sqrt{2}} & 0 & 0 \\
     0 & 0 & 0 & 0 & 0 & \frac{1}{\sqrt{2}} & -\frac{1}{\sqrt{2}} & 0 & 0 \\
     0 & 0 & 1 & 0 & 0 &0 & 0 & 0 & 0 \\
     0 & 0 & 0 &\frac{1}{\sqrt{2}}& 0 & 0 & 0 & \frac{1}{\sqrt{2}} & 0 \\
     0 & 0 & 0 &\frac{1}{\sqrt{2}}& 0 & 0 & 0 &  -\frac{1}{\sqrt{2}} &0 \\
   \end{array}
\right).
\end{eqnarray}
The orthogonal matrix $P$ satisfies the relation
$PP^{T}=P^{T}P=1_{9\times9}$, where $P^{T}$ denotes the transpose of
matrix $P$.

By means of $P$, the Hamiltonian $\hat{H}$ for
$\varphi_{1}=\varphi_{2}$ can be recast as follows,
\begin{eqnarray}\label{12}
\tilde{\hat{H}}&=&P\hat{H}P^{T}\nonumber\\
&=&diag\{H^{(1)}_{\frac{1}{2}},H^{(1)}_{0},H^{(2)}_{0},H^{(2)}_{\frac{1}{2}},H^{(3)}_{0},H^{(3)}_{\frac{1}{2}}\}\\
&=&\bigoplus_{k=1}^{3} \tilde{H}^{(k)},\nonumber
\end{eqnarray}
where $\tilde{H}^{(k)}=H^{(k)}_{\frac{1}{2}}\oplus H^{(k)}_{0}$,
$H^{(k)}_{\frac{1}{2}}$'s are $2\times2$ matrices, and
$H^{(k)}_{0}$'s are $1\times1$ matrices with $H^{(k)}_{0}=(0)$. The
$(3\times3)$-dimensional interactional Hamiltonian system is
decomposed into six subsystems. Three sets of $SU(2)$ realizations
(\ref{rlz1}) can be recast as,
\begin{eqnarray}\label{10}
 \tilde{S}^{(1)}_{+}=|00\rangle\langle 01|,~~~~
\tilde{S}^{(1)}_{-}=|01\rangle\langle 00|,~~~~
\tilde{S}^{(1)}_{3}=\frac{1}{2}(|00\rangle\langle
00|-|01\rangle\langle 01|),
\end{eqnarray}
\begin{eqnarray}
 \tilde{S}^{(2)}_{+}=|11\rangle\langle 20|,~~~~
\tilde{S}^{(2)}_{-}=|20\rangle\langle 11|,~~~~
\tilde{S}^{(2)}_{3}=\frac{1}{2}(|11\rangle\langle
11|-|20\rangle\langle 20|),
\end{eqnarray}
\begin{eqnarray}
 \tilde{S}^{(3)}_{+}=|21\rangle\langle 22|,~~~~
\tilde{S}^{(3)}_{-}=|22\rangle\langle 21|,~~~~
\tilde{S}^{(3)}_{3}=\frac{1}{2}(|21\rangle\langle
21|-|22\rangle\langle 22|),
\end{eqnarray}
 and
the seconde-order Casimir operators are $\tilde{\mathcal
{J}}^{(1)}=\frac{3}{4}(|00\rangle\langle 00|+|01\rangle\langle
01|)$, $\tilde{\mathcal {J}}^{(2)}=\frac{3}{4}(|11\rangle\langle
11|+|20\rangle\langle 20|)$, $\tilde{\mathcal
{J}}^{(3)}=\frac{3}{4}(|21\rangle\langle 21|+|22\rangle\langle
22|)$.


\baselineskip 22pt

\begin{thebibliography}{99}


\bibitem{berry} M. V. Berry, Proc. R. Soc. London,  {\it Ser. A}  {\bf 392} (1984) 45.
\bibitem{BS}B. Simon, {\it Phys. Rev. Lett.} {\bf 51} (1983) 2167--2170.
\bibitem{aa} Y. Aharonov and J. Anandan, {\it Phys. Rev. Lett.} {\bf 58} (1987) 1593.

\bibitem{Esj} E. Sj\"{o}vist, A. K. Pati, A. Ekert, J. S. Anandan, M. Ericsson, D. K. L. Oi, and V. Vedral, {\it Phys. Rev. Lett.} {\bf 85} (2000)
2845.


\bibitem{JR} Samuel J and Bhandari R, {\it Phys. Rev. Lett.} {\bf
60} (1988) 2339.

\bibitem{DELC} Tong D M, Sj\"{o}qvist E, Kwek L C, and Oh C H, {\it Phys. Rev.
Lett.} {\bf 93} (2004) 080405.

\bibitem{FA} Wilczek F and Zee A, {\it Phys. Rev. Lett.} {\bf 52} (1984) 2111.

\bibitem{DGRA} Suter D, Chingas G, Harris R and Pines A, {\it Molec. Phys.} {\bf
61} (1987) 1327.
\bibitem{MVM} Goldman M, Fleury V and Gu$\acute{e}$ron M, {\it J. Magn.
Reson.} A {\bf 118} (1996) 11.
\bibitem{RT} Tycko R, {\it Phys. Rev. Lett.} {\bf 58} (1987) 228t -2284.
\bibitem{SGM} Appelt S, $\ddot{W}$ackerle G and Mehring M, {\it Phys. Rev. Lett.} {\bf
72} (1994) 3921.
\bibitem{JA} Jones J A and Pines A, {\it J. Chem. Phys.} {\bf 106} (1997)
3007.
\bibitem{RY} Chiao R Y and Wu Y S, {\it Phys. Rev. Lett.} {\bf 57} (1986) 8.

\bibitem{AAHQJZ} Bohm A, Mostafazadeh A, Koizumi H, Niu Q, Appelt J, Wackerle G and Mehring M, {\it Phys. Rev.
Lett.} {\bf 72} (1994) 3921.

\bibitem{J. Jones} Jones J, Vedral V, Ekert A K and Castagnoli C, {\it Nature} {\bf 403} (2000) 869; Duan L M, Cirac J I and Zoller P, {\it
Science} {\bf 292} (2001) 1695.

\bibitem{W. K.} Wootters W K, {\it Phys. Rev. Lett.} {\bf 80} (1998) 2245.

\bibitem{A. Ekert} Ekert A, Ericsson M, Hayden P, Inamori H, Jones J A, Oi D K L and Vedral V, {\it J. Mod. Opt.} {\bf 47} (2000) 2501.


\bibitem{Zanardi} Zanardi P and Rasetti M, {\it Phys. Lett. A.} {\bf 264} (1999) 94.

\bibitem{Bennett} Bennett C H and DiVincenzo D P, {\it Nature} {\bf 404} (2000) 247.
\bibitem{ben1} Bennett C H, Brassard G, Cr\'{e}peau C, Jozsa R, Peres A, and Wootters W K, {\it Phys. Rev. Lett.} {\bf 70} (1993) 1895.
\bibitem{ben2} Bennett C H and Wiesner S J, {\it Phys. Rev. Lett.} {\bf 69} (1992) 2881.
\bibitem{mdmv} Murao M, Jonathan D, Plenio M B, and Vedral V, {\it Phys. Rev. A} {\bf 59} (1999) 156.
\bibitem{yang} Yang C N: {\it Phys. Rev. Lett.} {\bf 19} (1967) 1312; Yang C N, {\it Phys. Rev.} {\bf
168} (1968) 1920.
\bibitem{baxter} Baxter R J, {\it Exactly Solved Models in Statistical Mechanics} (New York:
Academic) (1982);\\ Baxter R J, {\it Ann. Phys.} {\bf 70} (1972)
193.
\bibitem{drin} Drinfeld V G, {\it Soviet Math. Dokl} {\bf 32} (1985) 254; Drinfeld V G, {\it Soviet Math. Dokl} {\bf 36} (1988)
212; Drinfeld V G, {\it Quantum Groups (in Proc. ICM)} vol 269
(Berkeley, CA:Academic) (1986).
\bibitem{qiybe1} Kitaev A Y, {\it Ann. Phys.} {\bf 303} (2003) 2.
\bibitem{kauffman1} Kauffman L H and Lomonaco(Jr) S J, {\it New J. Phys.} {\bf
6} (2004) 134.
\bibitem{qiybe3} Franko J M, Rowell E C, and Wang Z, {\it J. Knot Theory Ramif.} {\bf 15} (2006) 413.
\bibitem{zkg} Zhang Y, Kauffman L H, and Ge M L, {\it Int. J. Quant. Inf.} {\bf
3} (2005) 669.
\bibitem{zg} Zhang Y and Ge M L, {\it Quant. Inf. Proc.} {\bf
6} (2007) 363, Zhang Y, Rowell E C, Wu Y S, Wang Z H and Ge M L,
{\it e-print} quant-ph/0706.1761.
\bibitem{ckg} Chen J L, Xue K, and Ge M L, {\it Phys. Rev. A.} {\bf 76} (2007)
042324.
\bibitem{ckg2} Chen J L, Xue K, and Ge M L: {\it Ann. Phys.} {\bf
323} (2008) 2614.
\bibitem{cxg1} Chen J L, Xue K, and Ge M L, {\it e-print} quant-ph/0809.2321.
\bibitem{S. W. Hu}Hu S W, Xue K, and Ge M L, {\it Phys. Rev. A} {\bf
78} (2008) 022319.
\bibitem{wang1} Wang G C, Xue K, Wu C F, Liang H and Oh C H, {\it J. Phys. A: Math. Theor.}
\textbf{42} (2009) 125207.
\bibitem{Zyczkowski} Zyczkowski K, Horodecki P,
sanpera A, and lewenstein M, {\it Phys. Rev. A} {\bf 58} (1998) 883,
Appendix: Zyczkowski B K has used the same quantity in several other
works to characterize the entanglement of mixed states. See, for
instance: {\it Phys Rev. A} {\bf 60} (1999) 3496.
\bibitem{G. Vidal} Vidal G and Werner R F, {\it Phys. Rev. A} {\bf
65} (2002) 032314.
\bibitem{smv} Zhang W, Feng D and Gilmore R, {\it Rev. Mod. Phys.} {\bf
62} (1990) 867; Chaturvedi S, Sriram M S and Srinivasan V, {\it J.
Phys. A: Math. Gen} {\bf 20} (1987) L1091.
\end{thebibliography}
\end{document}